\documentclass[%
 reprint,
 amsmath,amssymb,
 aps,pra,onecolumn,superscriptaddress,floatfix
]{revtex4-2}

\usepackage{graphicx}
\usepackage{dcolumn}
\usepackage{bm}
\usepackage{qcircuit}
\usepackage{tikz}
\usetikzlibrary{arrows, positioning, calc, shapes.geometric, fit}
\usepackage{booktabs}
\usepackage{amsthm}
\usepackage{physics}
\usepackage{xcolor}
\usepackage{multirow}

\usepackage{hyperref}

\graphicspath{{figures/}}

\hypersetup{
    colorlinks=true,
    linkcolor=blue,
    citecolor=blue,
    urlcolor=blue
}

\newcommand{\cE}{\mathcal{E}}

\newcommand{\cL}{\mathcal{L}}
\newcommand{\id}{\mathbb{I}}
\renewcommand{\tr}{\operatorname{Tr}}

\newcommand{\CPTP}{\mathrm{CPTP}}

\newtheorem{proposition}{Proposition}

\begin{document}

\title{Observable-Level Expressivity in Fixed-Readout Quantum Neural Networks}

\author{Iqra Bashir}
\affiliation{College of Physics and Optoelectronic Engineering, Shenzhen University, Shenzhen 518060, China}

\author{Yuanfeng Wang}
\affiliation{Quantum Science Center of Guangdong-Hong Kong-Macao Greater Bay Area, Shenzhen 518045, China}

\author{Mingda Huang}
\affiliation{Quantum Science Center of Guangdong-Hong Kong-Macao Greater Bay Area, Shenzhen 518045, China}

\author{Keren Li}
\email{likr@szu.edu.cn}
\affiliation{College of Physics and Optoelectronic Engineering, Shenzhen University, Shenzhen 518060, China}
\affiliation{Quantum Science Center of Guangdong-Hong Kong-Macao Greater Bay Area, Shenzhen 518045, China}

\author{Qi Qin}
\email{qiqin32@cityu.edu.hk}
\affiliation{Department of System Engineering, City University of Hong Kong, 83 Tat Chee Avenue, Kowloon, Hong Kong SAR, China}
\affiliation{Quantum Science Center of Guangdong-Hong Kong-Macao Greater Bay Area, Shenzhen 518045, China}

\begin{abstract}
A central question in quantum machine learning is which functions of an input state a model can represent when its measurement is fixed by hardware or by experimental design.
We address this question for quantum neural networks with a fixed readout observable, where the trainable pre-measurement dynamics determine the effective observable in the Heisenberg picture and the scalar output is the measured expectation value itself, without trainable affine post-processing.
We characterize the reachable effective observables under three physical resource classes: unitary evolution, which preserves the spectrum of the readout; mixed-unitary evolution, which relaxes this to a majorization constraint while preserving the trace; and completely positive trace-preserving (CPTP) evolution, whose Heisenberg adjoint is unital rather than trace-preserving and therefore lifts the trace constraint inherited at the mixed-unitary level.
The three reachable sets are a fixed-spectrum orbit, its convex hull, and an operator interval, and the inclusions are strict for any readout that is not proportional to the identity.
The separation is witnessed locally: for a fixed Pauli readout $Z_j$ on an arbitrary number of qubits, the amplitude-damping observable $\gamma\id+(1-\gamma)Z_j$ lies outside the mixed-unitary level for every $\gamma>0$.
We realize the CPTP level with a Stinespring-channelized circuit and derive closed-form loss floors, $4\gamma^2/3$ and $\gamma^2$ for the unitary and mixed-unitary levels and zero at the CPTP level, which the numerics reproduce within precision.
Structured-observable, R\'enyi-moment and classification experiments illustrate how the hierarchy manifests in finite-depth ans\"atze.
The results identify which spectral constraint each physical resource relaxes, framing reachable-observable classes as a resource for expressivity rather than an architectural advantage.
\end{abstract}
\maketitle

\section{Introduction}

A fundamental question in quantum learning theory is what scalar functions of quantum states can be represented by parameterized quantum circuits.
This expressivity question has been studied from multiple perspectives, including circuit expressibility and entangling capability~\cite{sim2019expressibility,du2022efficient}, Fourier spectra of data-encoding circuits~\cite{schuld2021effect,perez2020data}, effective dimension~\cite{Abbas2021The, yao2026learningmaximizequantumneural}, learnability and trainability~\cite{du2021learnability,Caro2022,Banchi2021,li2026learning}, and barren-plateau behavior~\cite{mcclean2018barren,Cerezo2021Cost,Holmes2022Connecting}.
Parameterized circuits in this context are often termed quantum neural networks (QNNs) and applied to quantum state classification~\cite{sharma2025quantumn,sander2024quantum}, Hamiltonian and molecular-energy learning~\cite{wang2025qghnn,li2025quantum}, dynamical modeling~\cite{Choi2022Learning}, quantum autoencoding~\cite{Romero2017Quantum}, generative modeling~\cite{tian2022recent}, and variational simulation~\cite{Zhang2022Variational}.

In many learning scenarios the measurement apparatus is constrained by the available measurement primitive, by tomographic overhead, or by hardware limitations, while the quantum dynamics applied before measurement remain trainable.
This regime is physically motivated rather than a mathematical simplification: implementing an arbitrary target observable requires either a hardware-native measurement basis unavailable on near-term devices or additional tomographic overhead to reconstruct it, whereas the pre-measurement dynamics can be varied freely within the trainable circuit.
Under this constraint the trainable dynamics fully determine the effective observable, and it is this observable -- not the physical readout -- that fixes the model's hypothesis class.
Existing expressivity frameworks address neighbouring but distinct objects: state-space coverage and entangling capability~\cite{sim2019expressibility,du2022efficient}, the frequency content of data-encoding circuits~\cite{schuld2021effect,perez2020data}, and model capacity~\cite{Abbas2021The}.
None of them characterizes which observables are reachable once the readout is fixed, which is the quantity that determines the representable functions in this regime.
This fixed-readout setting therefore calls for an observable-level notion of expressivity, in which the scalar output is taken to be the measured expectation value itself, without additional trainable affine post-processing.
If a trainable unitary $U(\theta)$ precedes a fixed Hermitian readout $M$, the prediction functional becomes
\begin{equation}
    f_\theta(\rho)=\tr[M U(\theta)\rho U^\dagger(\theta)]
    =\tr[M_{\mathrm{eff}}(\theta)\rho],
\end{equation}
where the effective observable in the Heisenberg picture is
\begin{equation}
    M_{\mathrm{eff}}(\theta)=U^\dagger(\theta)M U(\theta).
    \label{eq:intro_unitary_orbit}
\end{equation}
The hypothesis class is thus determined by which effective observables $M_{\mathrm{eff}}$ are reachable: writing $\mathcal{R}$ for the reachable set under a given class of pre-measurement dynamics, the induced hypothesis class is
\begin{equation}
    \mathcal{F}_{\mathcal{R}} = \left\{\, \rho \mapsto \tr[O\rho] \;:\; O \in \mathcal{R} \,\right\},
    \label{eq:hypothesis_class}
\end{equation}
so that characterizing $\mathcal{R}$ is equivalent to characterizing the functions the model can represent.
This viewpoint is closely related to recent analyses of trainable observables and measurement design~\cite{wu2021expressibility,lin2025adaptive,yang2025measuring}, but emphasizes the spectral and structural constraints imposed by the fixed readout.
A key bottleneck emerges: increasing circuit depth may improve coverage of the unitary orbit $\{U^\dagger M U\}$, but cannot alter spectral invariants such as the eigenvalues or trace of $M$.

This work addresses the following question: \emph{given a fixed readout observable $M$, which effective observables can be generated when the trainable dynamics are subject to different physical constraints?}
We consider three physically motivated classes, corresponding to three distinct resources.
\begin{itemize}
    \item \textbf{Unitary evolution} (coherence): $M_{\mathrm{eff}}=U^\dagger M U$. Closed-system dynamics preserve the full spectrum of the readout.
    \item \textbf{Mixed-unitary evolution} (classical randomness): $M_{\mathrm{eff}}=\sum_i p_i U_i^\dagger M U_i$. Averaging over unitaries relaxes the spectral constraint to a majorization relation~\cite{nielsen2002majorization,Uhlmann1971,Watrous2018}, but every term in the convex sum has the trace of $M$, so the trace is preserved.
    \item \textbf{CPTP evolution} (environment coupling): $M_{\mathrm{eff}}=\cE^\dagger(M)$ for a completely positive trace-preserving map $\cE$. The channel remains trace-preserving on states; its Heisenberg adjoint $\cE^\dagger$ is completely positive and \emph{unital} rather than trace-preserving. The trace of the effective observable is then no longer tied to that of $M$, which lifts the constraint inherited at the mixed-unitary level and admits non-unital effective observables.
\end{itemize}
These classes form a hierarchy of reachable sets under progressively weaker spectral constraints.
Our main contribution is a characterization of these sets and a proof that the hierarchy is strict, for any readout that is not proportional to the identity, with separations detectable by local single-qubit witnesses at arbitrary system size.

A minimal example shows that the distinction between these classes follows from spectral structure rather than from circuit optimization.
For a one-qubit readout $M=Z$, every mixed-unitary observable $\sum_i p_i U_i^\dagger Z U_i$ is traceless, since unitary conjugation preserves the trace and $\tr Z=0$; no convex combination of traceless operators carries an identity component.
Amplitude damping with strength $\gamma\in[0,1]$~\cite{nielsen2010quantum,Watrous2018} induces the effective observable $\cE_{\mathrm{AD}}^\dagger(Z)=\gamma\id+(1-\gamma)Z$, with $\tr[\gamma\id]=2\gamma\neq0$ for $\gamma>0$.
The obstruction is an invariant of the reachable set, not a property of any particular ansatz or optimizer: no circuit depth, parameter count or training schedule can produce an identity component from mixed-unitary resources.
The same obstruction persists when this local readout is embedded in an arbitrary many-qubit system, which makes it a scalable witness of the separation.

To formalize these distinctions, we define reachable-observable sets for each physical class and establish their inclusion structure: the unitary orbit is contained in its convex hull (the mixed-unitary set), which is in turn contained in the CPTP-induced set.
The formal definitions and inclusion proofs are given in Sec.~\ref{sec:reachability}.
We then prove that both inclusions are strict whenever the readout has more than one distinct eigenvalue, by constructing explicit local witnesses based on spectral contraction and identity mixing.
The witness construction is presented in Sec.~\ref{sec:amplitude_damping}.

The CPTP-induced reachable class can be realized at the circuit level via Stinespring dilation.
A Stinespring-channelized QNN (SQNN) couples the input system to an ancilla register, applies a trainable joint unitary, discards the ancilla, and measures the fixed readout on the system.
In the Heisenberg picture, the fixed readout $M$ becomes
\begin{equation}
    M_{\mathrm{eff}}(\theta)
    =
    \cE_\theta^\dagger(M)
    =
    \sum_\mu K_\mu^\dagger(\theta)M K_\mu(\theta),
    \label{eq:intro_channelized_readout}
\end{equation}
where $K_\mu(\theta)$ are Kraus operators of the induced CPTP map $\cE_\theta$.
This architecture realizes the CPTP-induced reachable class when the ancilla dimension and unitary expressiveness are sufficiently large, and reduces to the mixed-unitary and unitary levels when the channel is restricted to mixed-unitary or single-unitary form.
The SQNN architecture and its limiting cases are described in Sec.~\ref{sec:qchann_architecture}.

Beyond architectural realization, we establish that the hierarchy can be separated using local single-qubit measurements for arbitrary system size.
For a local Pauli readout $Z_j$ on qubit $j$, Pauli-weight contraction distinguishes the unitary level from the mixed-unitary level, while identity-component mixing distinguishes the CPTP level from the mixed-unitary level.
These local witnesses demonstrate that the reachable-set separations are not artifacts of global entangling structure or many-body readouts, but follow from fundamental constraints on single-qubit observables.
The local-witness construction and separation proofs are presented in Sec.~\ref{sec:amplitude_damping}.

The contributions are as follows.
First, we give an exact characterization of the three reachable-observable sets for a fixed readout: a fixed-spectrum orbit, its convex hull described by majorization, and the operator interval $[m_{\min}\id,m_{\max}\id]$ (Proposition~\ref{prop:three_level_geometry}).
Second, we prove strictness of both inclusions with an explicit single-qubit witness family that applies at arbitrary system size and requires only a local Pauli readout (Proposition~\ref{prop:local_single_qubit_separation}).
Third, we convert the separation into closed-form loss floors for an observable-learning task, Eq.~\eqref{eq:three_level_loss_floors}, giving quantitative and falsifiable predictions rather than qualitative statements.
Fourth, we identify which physical resource relaxes which constraint: coherence preserves the spectrum, classical randomness relaxes it to majorization, and environment coupling lifts the trace constraint through the unitality of the Heisenberg adjoint.

The remainder of the paper is organized as follows.
Section~\ref{sec:theory} develops the reachable-observable hierarchy, proves the separations, and gives the Stinespring realization of the CPTP level.
Section~\ref{sec:observable_learning} reports the numerical work in two clearly separated tiers: Sec.~\ref{subsec:certified_separation} certifies the analytic loss floors of Eq.~\eqref{eq:three_level_loss_floors} against finite parametrizations whose optima are known in closed form, while Secs.~\ref{subsec:representative_benchmarks} and~\ref{subsec:trainability_control} give representative finite-ansatz evidence in structured-observable, R\'enyi-moment and classification settings.
Section~\ref{sec:discussion} sets out the scope of the claims, their limitations, and the relation to existing architectures and expressivity measures.
Implementation details are collected in the appendices.

\section{Theory}
\label{sec:theory}

\subsection{Reachable-Observable Hierarchy}
\label{sec:reachability}

The expressivity of a fixed-readout quantum neural network is determined by which effective observables can be generated from a given measurement operator through trainable pre-measurement dynamics. Different physical resources—closed-system evolution, classical randomness, or coupling to an environment—impose different constraints on this reachable set.

We formalize this structure as follows. Let $\mathcal{H}\simeq\mathbb{C}^d$ be the system Hilbert space, and let $M\in\mathrm{Herm}(\mathcal{H})$ be the fixed readout observable. Any trainable pre-measurement dynamics maps this readout, in the Heisenberg picture, to an effective observable $O\in\mathrm{Herm}(\mathcal{H})$. The model output is the linear functional
\begin{equation}
    f_O(\rho)=\mathrm{tr}[O\rho],
    \label{eq:target_functional}
\end{equation}
evaluated on the input state $\rho$.
Throughout this work, a fixed-readout model produces the measured expectation value itself as its scalar output, without trainable affine post-processing unless explicitly stated.

We compare three classes of pre-measurement dynamics, corresponding to different physical resources:
\begin{align}
    \mathcal{R}_U(M)
    &=
    \{U^\dagger M U:U\in\mathcal{U}(d)\},
    \label{eq:unitary_orbit_set}\\
    \mathcal{R}_{\mathrm{conv}}(M)
    &=
    \mathrm{conv}\, \mathcal{R}_U(M),
    \label{eq:conv_set}\\
    \mathcal{R}_{\mathrm{CPTP}}(M)
    &=
    \{\mathcal{E}^\dagger(M):\mathcal{E}\text{ is CPTP}\}.
    \label{eq:cptp_set}
\end{align}
Here $\mathcal{R}_U(M)$ corresponds to closed-system unitary dynamics, $\mathcal{R}_{\mathrm{conv}}(M)$ to statistical mixtures of unitary transformations, implementable through classical randomness (the mixed-unitary, or mixed-unitary, level), and $\mathcal{R}_{\mathrm{CPTP}}(M)$ to general quantum channels (physically realized through ancilla-mediated evolution).
An element of $\mathcal{R}_{\mathrm{conv}}(M)$ has the form $\sum_i p_i U_i^\dagger M U_i$ with $p_i\ge0$ and $\sum_i p_i=1$.
For a CPTP map $\mathcal{E}(\rho)=\sum_\mu K_\mu\rho K_\mu^\dagger$, the Heisenberg adjoint $\mathcal{E}^\dagger$, defined by $\mathrm{tr}[A\mathcal{E}(\rho)]=\mathrm{tr}[\mathcal{E}^\dagger(A)\rho]$, is completely positive and unital, and acts on the readout as
\begin{equation}
    \mathcal{E}^\dagger(M)=\sum_\mu K_\mu^\dagger M K_\mu,
    \quad
    \sum_\mu K_\mu^\dagger K_\mu=\mathbb{I}.
\end{equation}

The asymmetry between $\cE$ and $\cE^\dagger$ is the single structural fact behind the hierarchy.
A CPTP map is positive and trace-preserving on states; its adjoint is positive and unital on observables.
Unitality fixes $\cE^\dagger(\id)=\id$ and therefore confines the effective observable to the operator interval spanned by the spectrum of $M$, but it places no constraint on $\tr\,\cE^\dagger(M)$.
Mixed-unitary dynamics, whose adjoints are both unital and trace-preserving, retain the stronger constraint.

\begin{figure}[tbp]
\centering
\begin{tikzpicture}[
    font=\normalsize,
    >=stealth,
    setboundary/.style={draw, thick, ellipse, align=center},
    witness/.style={circle, fill=black, inner sep=1.2pt}
]
\path[use as bounding box] (-3.45,-1.45) rectangle (3.45,1.65);
\node[font=\bfseries] at (-3.35,1.55) {(a)};
\node[setboundary, fill=green!9, minimum width=6.35cm, minimum height=2.75cm] (cptp) at (0,0) {};
\node[setboundary, fill=orange!15, minimum width=4.35cm, minimum height=1.78cm] (convex) at (-0.35,0) {};
\node[setboundary, fill=blue!12, minimum width=2.15cm, minimum height=0.86cm] (unitary) at (-0.75,0) {};
\node at (-0.72,0.20) {$\mathcal{R}_U$};
\node at (-0.10,0.74) {$\mathcal{R}_{\mathrm{conv}}$};
\node at (1.50,1.08) {$\mathcal{R}_{\mathrm{CPTP}}$};
\node[witness] at (-0.78,-0.12) {};
\node[anchor=north] at (-0.78,-0.31) {$Z$};
\node[witness] at (0.68,-0.14) {};
\node[anchor=north] at (0.68,-0.33) {$rZ$};
\node[witness] at (1.82,-0.36) {};
\node[anchor=north, align=center] at (1.66,-0.55) {$O_\gamma$};
\end{tikzpicture}

\vspace{0.7em}

\begin{tikzpicture}[
    font=\normalsize,
    >=stealth,
    node distance=0.45cm,
    process/.style={draw, thick, rounded corners, fill=green!9, minimum height=0.72cm, align=center},
    readout/.style={draw, thick, rounded corners, fill=blue!12, minimum height=0.72cm, align=center},
    arrow/.style={->, thick}
]
\path[use as bounding box] (-3.45,-0.35) rectangle (3.55,1.65);
\node[font=\bfseries] at (-3.35,1.55) {(b)};
\node (rho) at (-3.00,0) {$\rho$};
\node (anc) at (-3.00,0.85) {$\ket{0}_a$};
\node[process, minimum width=1.7cm, minimum height=1.25cm] (U) at (-1.15,0.42) {$U_\theta^{sa}$};
\node[process, fill=orange!15, minimum width=1.05cm] (trace) at (0.80,0.42) {$\mathrm{tr}_a$};
\node[readout, minimum width=1.05cm] (meas) at (2.20,0.42) {$M$};
\node[anchor=west] (out) at (3.10,0.42) {$f_\theta(\rho)$};
\draw[arrow] (rho) -- (U.west |- rho);
\draw[arrow] (anc) -- (U.west |- anc);
\draw[arrow] (U.east) -- (trace.west);
\draw[arrow] (trace.east) -- (meas.west);
\draw[arrow] (meas.east) -- (3.02,0.42);
\end{tikzpicture}
\caption{Reachable-observable hierarchy and Stinespring realization. (a) Nested reachable sets for a fixed readout, with single-qubit witnesses: $Z\in\mathcal{R}_U(Z)$ (spectrum-preserving), $rZ\in\mathcal{R}_{\mathrm{conv}}(Z)\setminus\mathcal{R}_U(Z)$ for $0<|r|<1$ (majorized spectrum, trace-preserving), and $\gamma\mathbb{I}+(1-\gamma)Z\in\mathcal{R}_{\mathrm{CPTP}}(Z)\setminus\mathcal{R}_{\mathrm{conv}}(Z)$ for $\gamma>0$ (breaks trace preservation through dissipation). (b) Stinespring realization of the CPTP level: a system--ancilla unitary followed by ancilla tracing generates the effective observable $\mathcal{E}_\theta^\dagger(M)$ from the fixed readout $M$.}
\label{fig:general_qchann_architecture}
\end{figure}
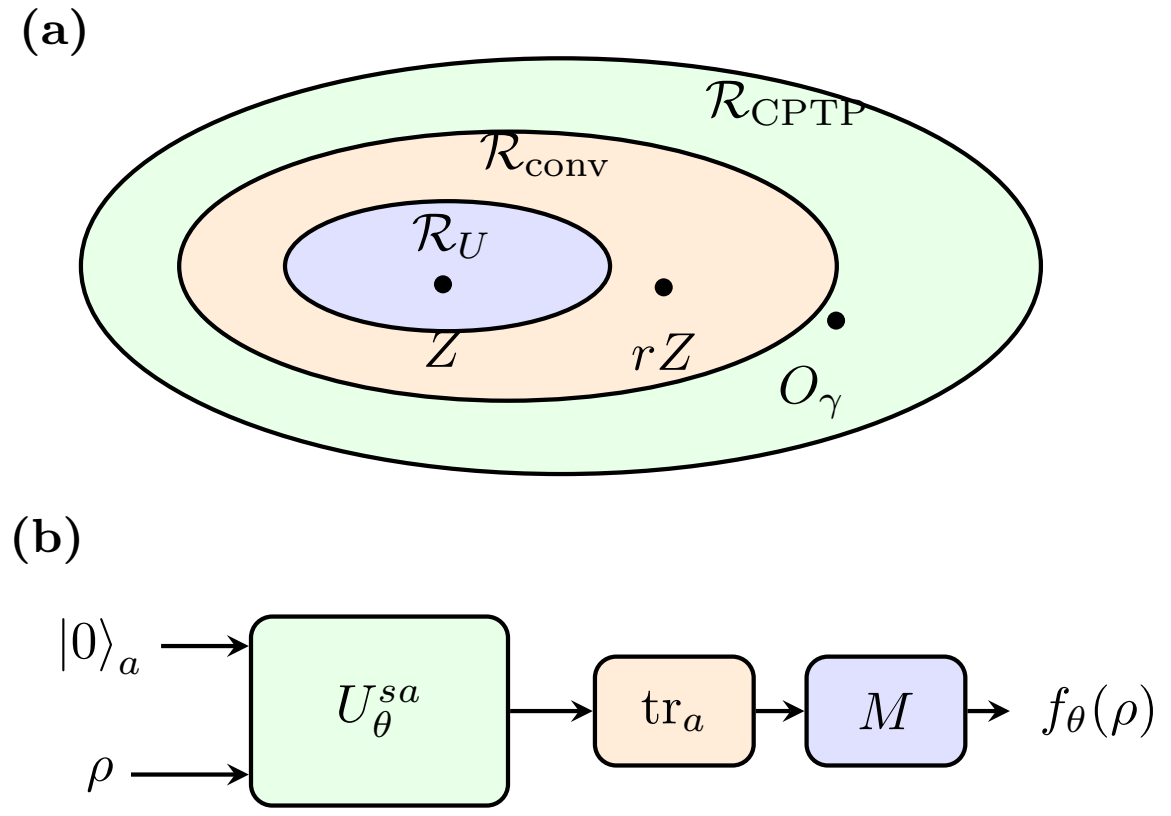

These three sets form a nested hierarchy, illustrated in Fig.~\ref{fig:general_qchann_architecture}(a).

\begin{proposition}
\label{prop:hierarchy}
For any fixed readout observable $M$, the reachable observable sets satisfy
\begin{equation}
    \mathcal{R}_U(M)\subseteq \mathcal{R}_{\mathrm{conv}}(M)\subseteq \mathcal{R}_{\mathrm{CPTP}}(M).
    \label{eq:hierarchy_inclusions}
\end{equation}
\end{proposition}

\begin{proof}
The first inclusion follows because a unitary transformation is a trivial convex combination. For the second inclusion, take an arbitrary element
\(
    O=\sum_i p_i U_i^\dagger M U_i\in \mathcal{R}_{\mathrm{conv}}(M)
\)
with $p_i\ge0$ and $\sum_i p_i=1$. Define the mixed-unitary channel
\begin{equation}
    \Phi(\rho)=\sum_i p_i U_i\rho U_i^\dagger
\end{equation}
with Kraus operators $K_i=\sqrt{p_i}U_i$ satisfying $\sum_i K_i^\dagger K_i = \mathbb{I}$. Then
\begin{equation}
    \Phi^\dagger(M)=\sum_i p_i U_i^\dagger M U_i=O,
\end{equation}
proving $\mathcal{R}_{\mathrm{conv}}(M)\subseteq\mathcal{R}_{\mathrm{CPTP}}(M)$.
\end{proof}

The physical distinction between these levels is encoded in progressively relaxed spectral constraints.

\begin{proposition}
\label{prop:three_level_geometry}
Let $M\in\mathrm{Herm}(\mathcal{H})$ have eigenvalues in the interval $[m_{\min},m_{\max}]$. Then:
\begin{align}
    \mathcal{R}_U(M)
    &=
    \{O\in\mathrm{Herm}(\mathcal{H}):\lambda(O)=\lambda(M)\},
    \label{eq:unitary_geometry}\\
    \mathcal{R}_{\mathrm{conv}}(M)
    &=
    \{O\in\mathrm{Herm}(\mathcal{H}):\lambda(O)\prec\lambda(M)\},
    \label{eq:conv_geometry}\\
    \mathcal{R}_{\mathrm{CPTP}}(M)
    &=
    \{O\in\mathrm{Herm}(\mathcal{H}):m_{\min}\mathbb{I}\le O\le m_{\max}\mathbb{I}\}.
    \label{eq:cptp_interval}
\end{align}
Eigenvalue lists are sorted in nonincreasing order; $\lambda(O)\prec\lambda(M)$ denotes majorization.
\end{proposition}

Through Eq.~\eqref{eq:hypothesis_class}, each line of Proposition~\ref{prop:three_level_geometry} is a statement about a hypothesis class.
The unitary level gives functionals whose range over all states is fixed by $\lambda(M)$; the mixed-unitary level admits a contracted range but a fixed value on the maximally mixed state, since $\tr[O\,\id/d]=\tr M/d$; the CPTP level relaxes that value to anything consistent with the operator interval.
The constraint separating the top two levels is thus a constraint on the model output at a single state, which is what makes it detectable with a fixed local readout.

\begin{proof}
Equation~\eqref{eq:unitary_geometry} is the standard characterization of unitary orbits of Hermitian operators. Equation~\eqref{eq:conv_geometry} is the Schur--Horn--Uhlmann theorem~\cite{nielsen2002majorization,Uhlmann1971,Watrous2018}, which in particular implies trace preservation: $\mathrm{tr}\, O=\mathrm{tr}\, M$ for all $O\in\mathcal{R}_{\mathrm{conv}}(M)$.

For Eq.~\eqref{eq:cptp_interval}, if $O=\mathcal{E}^\dagger(M)$ with $\mathcal{E}$ CPTP, then $\mathcal{E}^\dagger$ is completely positive and unital, so $m_{\min}\mathbb{I}\le M\le m_{\max}\mathbb{I}$ implies $m_{\min}\mathbb{I}\le\mathcal{E}^\dagger(M)\le m_{\max}\mathbb{I}$.
Conversely, let $O=\sum_j o_j\ket{j}\!\bra{j}$ satisfy the interval condition, and write $M=\sum_a m_a\ket{m_a}\!\bra{m_a}$. For each eigenvalue $o_j\in[m_{\min},m_{\max}]$, choose probabilities $p(a|j)$ such that $\sum_a p(a|j)m_a=o_j$. Define the POVM $\{F_a\}$ with $F_a=\sum_j p(a|j)\ket{j}\!\bra{j}\ge0$ and $\sum_a F_a=\mathbb{I}$. The measure-and-prepare channel
\begin{equation}
    \mathcal{E}(\rho)=\sum_a \mathrm{tr}(F_a\rho)\ket{m_a}\!\bra{m_a}
\end{equation}
satisfies $\mathcal{E}^\dagger(M)=\sum_a m_aF_a=O$, completing the characterization.
\end{proof}

Proposition~\ref{prop:three_level_geometry} reveals the physical meaning of each level. Unitary dynamics preserve the spectrum of $M$, corresponding to closed-system coherent evolution. Classical mixing relaxes this to majorization while preserving trace, allowing spectral redistribution through incoherent averaging. CPTP dynamics reach the full operator interval $[m_{\min}\id,m_{\max}\id]$.
The channel remains trace-preserving on states throughout; what is lifted is the constraint $\tr O=\tr M$ on the effective observable, because the Heisenberg adjoint $\cE^\dagger$ is unital rather than trace-preserving.
Environment coupling is the physical resource that removes this constraint.

For a Pauli readout $M=Z$, the CPTP-reachable set is the operator interval $-\mathbb{I}\le O\le\mathbb{I}$, whereas the mixed-unitary level is restricted to the trace-zero slice $\{\alpha\bm{\sigma}:\|\bm{\alpha}\|_2\le1\}$. Thus the amplitude-damping observable $O_\gamma=\gamma\id+(1-\gamma)Z$ lies in $\mathcal{R}_{\mathrm{CPTP}}(Z)$ but outside $\mathcal{R}_{\mathrm{conv}}(Z)$ for any $\gamma>0$: its identity component records the population transferred to the environment, and no trace-preserving observable transformation can generate it.

\subsection{Stinespring Realization of the CPTP Level}
\label{sec:qchann_architecture}

We now establish the connection between the reachable-set hierarchy and quantum circuit architectures. The CPTP level is physically realized through the Stinespring dilation: append an ancilla initialized to $\ket{0}_a$, apply a trainable system--ancilla unitary $U_\theta^{sa}$, and trace out the ancilla. This produces the system channel
\begin{equation}
    \mathcal{E}_\theta(\rho)
    =
    \mathrm{tr}_a\!\left[
    U_\theta^{sa}(\rho\otimes\ket{0}_a\!\bra{0}_a)U_\theta^{sa\dagger}
    \right],
    \label{eq:stinespring_channel}
\end{equation}
with Kraus representation
\begin{equation}
    \mathcal{E}_\theta(\rho)
    =
    \sum_\mu K_\mu(\theta)\rho K_\mu^\dagger(\theta),
    \qquad
    K_\mu(\theta)={}_a\!\bra{\mu}U_\theta^{sa}\ket{0}_a,
    \label{eq:kraus_channel}
\end{equation}
where $\{\ket{\mu}_a\}$ is an ancilla basis and $\sum_\mu K_\mu^\dagger(\theta)K_\mu(\theta)=\mathbb{I}$. 
The workflow is shown in Fig.~\ref{fig:general_qchann_architecture}(b).

The fixed readout transforms under the Heisenberg adjoint of this channel:
\begin{equation}
    f_\theta(\rho)
    =
    \mathrm{tr}[M\mathcal{E}_\theta(\rho)]
    =
    \mathrm{tr}[\mathcal{E}_\theta^\dagger(M)\rho],
    \quad
    \mathcal{E}_\theta^\dagger(M)
    =
    \sum_\mu K_\mu^\dagger(\theta)M K_\mu(\theta).
    \label{eq:cptp_heisenberg_observable}
\end{equation}
The trainable effective observable is therefore $\mathcal{E}_\theta^\dagger(M)$, which lies in $\mathcal{R}_{\mathrm{CPTP}}(M)$ rather than being restricted to a unitary conjugate or convex combination of such conjugates.

By the Stinespring dilation theorem~\cite{stinespring1955positive,Watrous2018}, every CPTP map on $\mathcal{H}$ can be realized as
\begin{equation}
    \mathcal{E}(\rho)=\mathrm{tr}_a[V(\rho\otimes\ket{0}_a\!\bra{0}_a)V^\dagger]
\end{equation}
for some ancilla Hilbert space and system--ancilla unitary $V$. Consequently, with sufficient ancilla dimension and unrestricted joint unitaries, this construction can generate any element of $\mathcal{R}_{\mathrm{CPTP}}(M)$. Saturation requires unrestricted joint unitaries and sufficient ancilla dimension. At fixed depth, fixed ancilla number and a restricted gate set, the construction accesses a proper subset of $\mathcal{R}_{\mathrm{CPTP}}(M)$ in general, and the numerical models of Sec.~\ref{sec:observable_learning} should be read as representatives of a level rather than as exhaustive parametrizations of it. Characterizing the reachable subset as a function of circuit resources is left to future work.

The lower hierarchy levels appear as special cases. If the ancilla dimension is one (no ancilla), the construction reduces to system-only unitary dynamics: $\mathcal{E}_\theta^\dagger(M)=V_\theta^\dagger M V_\theta\in\mathcal{R}_U(M)$. If the joint unitary has the orthogonal-branch form $\sum_i\ket{i}\!\bra{i}_a\otimes V_i(\theta)$ and the ancilla is prepared in a superposition $\sum_i w_i(\phi)\ket{i}_a$ with $p_i(\phi)=|w_i(\phi)|^2$, tracing the ancilla yields the mixed-unitary channel
\begin{equation}
    \Phi_{\theta,\phi}(\rho)=\sum_i p_i(\phi)V_i(\theta)\rho V_i^\dagger(\theta),
\end{equation}
whose Heisenberg action generates $\mathcal{R}_{\mathrm{conv}}(M)$.

\subsection{Witness Observables and Observable-Learning Loss Floors}
\label{sec:amplitude_damping}

To separate the hierarchy experimentally, we construct explicit target observables that belong to one reachable level but not to the next lower one. We call such targets \emph{witness observables}. For a fixed Pauli readout, a single local qubit suffices.

Since any local Pauli readout $\bm{n}\cdot\bm{\sigma}_j$ is locally unitarily equivalent to $Z_j$, we consider the fixed readout $M=Z_j$ and the two-parameter family
\begin{equation}
    O_{\alpha,r}^{(j)}=\alpha\mathbb{I}+rZ_j.
    \label{eq:three_level_family}
\end{equation}

\begin{proposition}[Single-qubit witnesses]
\label{prop:local_single_qubit_separation}
For any $n$-qubit system with local readout $M=Z_j$, the family in Eq.~\eqref{eq:three_level_family} separates the three reachable levels:
\begin{align}
    &Z_j\in\mathcal{R}_U(Z_j),
    \nonumber\\
    &rZ_j\in\mathcal{R}_{\mathrm{conv}}(Z_j)\setminus\mathcal{R}_U(Z_j),
    \qquad 0<|r|<1,
    \nonumber\\
    &\alpha\mathbb{I}+rZ_j\in\mathcal{R}_{\mathrm{CPTP}}(Z_j)
    \setminus\mathcal{R}_{\mathrm{conv}}(Z_j),
    \quad
    \alpha\ne0,\ |\alpha|+|r|\le1.
    \label{eq:three_level_membership}
\end{align}
\end{proposition}

\begin{proof}
The first statement holds by choosing the identity unitary. For $0<|r|<1$, the observable $rZ_j$ has eigenvalues $\pm r$ rather than $\pm1$, so it does not lie in the spectrum-preserving set $\mathcal{R}_U(Z_j)$. However, it can be written as the convex combination
\begin{equation}
    rZ_j
    =
    \frac{1+r}{2}Z_j+
    \frac{1-r}{2}(-Z_j),
\end{equation}
where $-Z_j=X_jZ_jX_j\in\mathcal{R}_U(Z_j)$, proving $rZ_j\in\mathcal{R}_{\mathrm{conv}}(Z_j)$.

By Proposition~\ref{prop:three_level_geometry}, $\mathcal{R}_{\mathrm{CPTP}}(Z_j)=\{O:-\mathbb{I}\preceq O\preceq\mathbb{I}\}$. The eigenvalues of $O_{\alpha,r}^{(j)}$ are $\alpha\pm r$, so the condition $|\alpha|+|r|\le1$ is equivalent to membership in the CPTP interval. If $\alpha\ne0$, then $\mathrm{tr}\, O_{\alpha,r}^{(j)}=\alpha 2^n\ne0$, whereas trace preservation requires $\mathrm{tr}\, O=\mathrm{tr}\, Z_j=0$ for all $O\in\mathcal{R}_{\mathrm{conv}}(Z_j)$. Hence $O_{\alpha,r}^{(j)}$ lies in $\mathcal{R}_{\mathrm{CPTP}}(Z_j)\setminus\mathcal{R}_{\mathrm{conv}}(Z_j)$.
\end{proof}

The condition on the readout is sharp in the trivial direction: if $M=c\,\id$ then $\mathcal{R}_U(M)=\mathcal{R}_{\mathrm{conv}}(M)=\mathcal{R}_{\mathrm{CPTP}}(M)=\{c\,\id\}$ and the hierarchy collapses.
Any readout with at least two distinct eigenvalues admits the witness construction above after a local unitary change of basis.

A physically natural CPTP witness is provided by local amplitude damping. For a single qubit, the amplitude-damping channel with strength $\gamma\in[0,1]$ has Kraus operators~\cite{nielsen2010quantum,Watrous2018}
\begin{equation}
    K_0=
    \begin{pmatrix}
    1&0\\
    0&\sqrt{1-\gamma}
    \end{pmatrix},
    \qquad
    K_1=
    \begin{pmatrix}
    0&\sqrt{\gamma}\\
    0&0
    \end{pmatrix},
    \label{eq:amplitude_damping_kraus}
\end{equation}
whose Heisenberg action on the Pauli-$Z$ readout gives
\begin{equation}
    \mathcal{E}_\gamma^\dagger(Z)
    =
    \gamma\mathbb{I}+(1-\gamma)Z
    \equiv O_\gamma.
    \label{eq:amplitude_damping_observable}
\end{equation}
The identity term arises from population transfer to the environment: the channel drives excited states toward the ground state, and the resulting effective observable violates the trace constraint $\tr O=\tr Z=0$ that every element of $\mathcal{R}_{\mathrm{conv}}(Z)$ satisfies. Applied locally to qubit $j$ and as the identity channel elsewhere, this yields the witness
\begin{equation}
    \mathcal{E}_{\gamma,j}^\dagger(Z_j)
    =
    \gamma\mathbb{I}+(1-\gamma)Z_j,
    \label{eq:local_many_qubit_shift}
\end{equation}
which is the CPTP witness in Proposition~\ref{prop:local_single_qubit_separation} with $\alpha=\gamma$ and $r=1-\gamma$.

We quantify the separation through an observable-learning task. For Haar-random pure single-qubit input states, define the mean squared error
\begin{equation}
    \mathcal{L}(E;O_\gamma)
    =
    \mathbb{E}_\rho
    \left[
    \left(
    \mathrm{tr}[\rho(O_\gamma-E)]
    \right)^2
    \right].
    \label{eq:population_loss_gamma}
\end{equation}
The optimal estimator within each reachable level is found by minimizing over that level's parameterization. For the fixed readout $Z$, the unitary level corresponds to $E=\bm{r}\cdot\bm{\sigma}$ with $\|\bm{r}\|_2=1$, and the mixed-unitary level to $\|\bm{r}\|_2\le1$. Using the standard Haar average over pure states,
\begin{equation}
    \mathcal{L}(E;O_\gamma)
    =
    \gamma^2
    +
    \frac{\|(1-\gamma)\hat{\bm{z}}-\bm{r}\|_2^2}{3},
    \label{eq:bloch_loss_bound}
\end{equation}
which yields the optimal loss floors
\begin{equation}
\begin{aligned}
    \mathcal{L}_U^\star(O_\gamma)&=\frac{4\gamma^2}{3},
    \qquad
    \mathcal{L}_{\mathrm{conv}}^\star(O_\gamma)&=\gamma^2,
    \qquad
    \mathcal{L}_{\mathrm{CPTP}}^\star(O_\gamma)&=0.
\end{aligned}
\label{eq:three_level_loss_floors}
\end{equation}
The nonzero floor at the mixed-unitary level is the squared amplitude of the unreachable identity component, $\gamma^2$, and the additional $\gamma^2/3$ at the unitary level is the cost of the fixed spectrum. Both floors are properties of the reachable sets, so any model whose effective observable lies in the corresponding class inherits them regardless of parametrization, depth or optimizer. These analytic predictions provide quantitative benchmarks for the numerical experiments in Sec.~\ref{sec:observable_learning}; a single-ancilla circuit realization of amplitude damping is given in App.~\ref{app:amp_circuit}.

\section{Numerical Verification}
\label{sec:observable_learning}

This section numerically verifies predictions of the reachable-observable hierarchy developed in Sec.~\ref{sec:theory}. We organize the results into three parts. Section~\ref{subsec:certified_separation} provides certified verification of the local hierarchy separation using analytic loss floors derived in Sec.~\ref{sec:amplitude_damping}. Section~\ref{subsec:representative_benchmarks} presents representative finite-ansatz evidence of how spectral and trace constraints manifest in structured learning tasks. Section~\ref{subsec:trainability_control} reports a classification control experiment to confirm that CPTP ans\"atze remain trainable in settings where the hierarchy is not explicitly probed.
Throughout, Std-QNN, RE-QNN and SQNN denote finite parametrizations chosen as representatives of the unitary, mixed-unitary and CPTP levels. They are not proposed as competing architectures, and no comparison between them is intended as an architectural benchmark.

For all experiments in Secs.~\ref{subsec:certified_separation} and~\ref{subsec:representative_benchmarks}, labels are exact expectation values of a target observable $O_\star$ on Haar-random pure states,
\begin{equation}
    y_m=\bra{\psi_m}O_\star\ket{\psi_m},
    \qquad
    \ket{\psi_m}\sim\mathrm{Haar}.
    \label{eq:numerical_labels}
\end{equation}
For a learned effective observable $M_{\mathrm{eff}}$, we evaluate the empirical MSE
\begin{equation}
    \widehat{\cL}
    =
    \frac{1}{N_D}\sum_{m=1}^{N_D}
    \left(
    y_m-\bra{\psi_m}M_{\mathrm{eff}}\ket{\psi_m}
    \right)^2,
    \label{eq:empirical_mse}
\end{equation}
where $N_D$ is the number of states in the dataset. Unless otherwise stated, each experiment uses $512$ Haar-random training states and $4096$ test states, sampled in antithetic pairs as described in App.~\ref{app:numerics_details}. All compared models use identical datasets and evaluation procedures.

\subsection{Certified verification against analytic loss floors}
\label{subsec:certified_separation}

Proposition~\ref{prop:three_level_geometry} characterizes the three reachable levels for a fixed single-qubit readout as a fixed-spectrum orbit, its convex hull, and an operator interval, and Proposition~\ref{prop:local_single_qubit_separation} exhibits targets separating them. Equation~\eqref{eq:three_level_loss_floors} provides analytic optimal loss values for specific targets in the slice $O_{\alpha,r}=\alpha\id+rZ$ when $M=Z$. For the amplitude-damping witness $O_\gamma=\gamma\id+(1-\gamma)Z$, the predicted losses are $\mathcal{L}_U^\star=4\gamma^2/3$ (unitary), $\mathcal{L}_{\mathrm{conv}}^\star=\gamma^2$ (mixed-unitary), and $\mathcal{L}_{\mathrm{CPTP}}^\star=0$ (CPTP-induced).
The experiments in this subsection are certified in the following sense: for each hierarchy level the optimal achievable loss is known in closed form from Eq.~\eqref{eq:three_level_loss_floors}, and the finite parametrizations used here are chosen so that their optima coincide with those values (App.~\ref{app:numerics_details}). Agreement between the trained loss and the analytic floor therefore tests the characterization itself, not the quality of an ansatz.

We implement three finite-parametrization representatives of the hierarchy levels, chosen to realize the corresponding effective-observable classes: a standard unitary QNN (Std-QNN) with $M_{\mathrm{eff}}=U^\dagger ZU$; a randomized-ensemble QNN (RE-QNN) with $M_{\mathrm{eff}}=\sum_i p_iU_i^\dagger ZU_i$; and a Stinespring QNN (SQNN) with $M_{\mathrm{eff}}=\sum_\mu K_\mu^\dagger ZK_\mu$ using an amplitude-damping-capable ansatz. Concrete parametrizations are given in App.~\ref{app:numerics_details}. These representatives are used to test whether the numerical loss floors match the analytic predictions.

\begin{figure}[tbp]
\centering
\includegraphics[width=0.5\columnwidth]{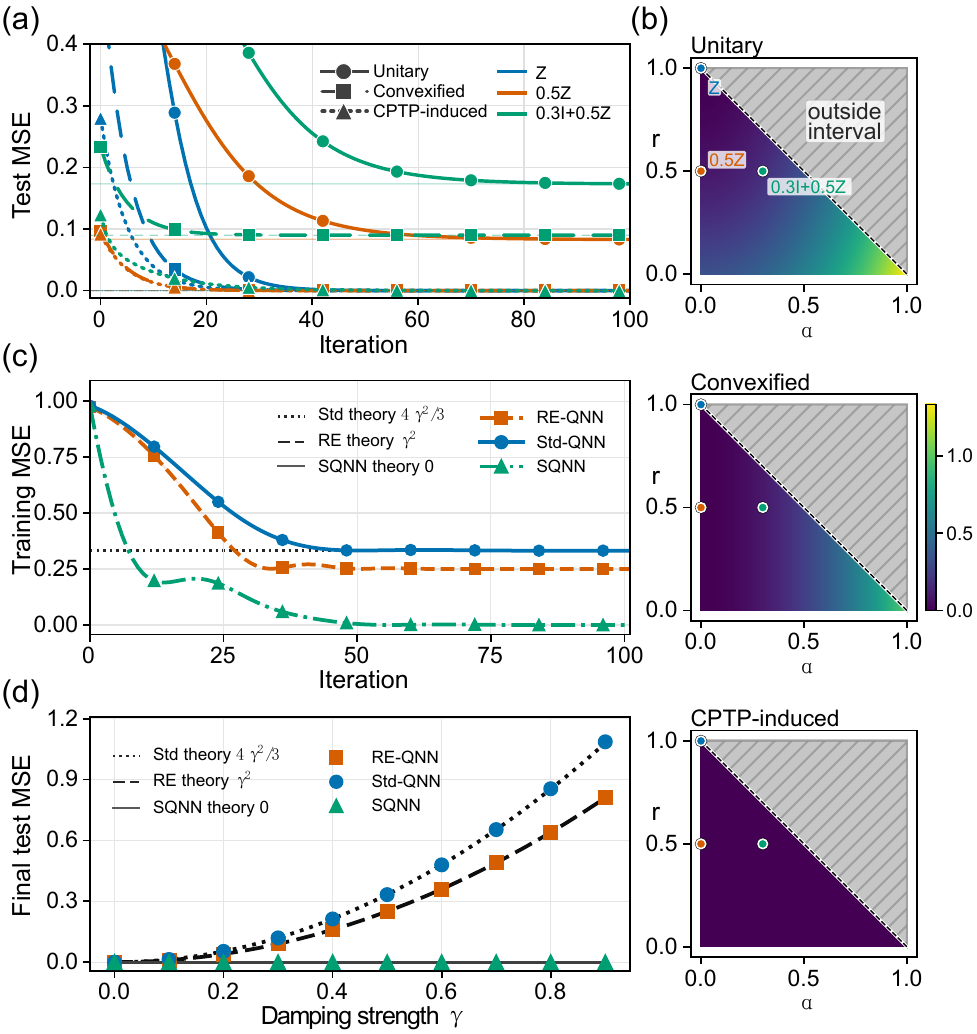}
\caption{Certified verification of the three-level reachable-observable hierarchy with fixed readout $M=Z$.
(a) Test MSE for targets in Eq.~\eqref{eq:three_level_targets}; horizontal lines mark analytic optimal losses.
(b) Analytic optimal loss over the slice $O_{\alpha,r}=\alpha\id+rZ$ for the three reachable levels. The dashed line $\alpha+r=1$ bounds the CPTP-induced operator interval.
(c) Training curves for the amplitude-damping witness at $\gamma=0.5$.
(d) Final test MSE versus $\gamma$; curves show analytic predictions $\cL_U^\star=4\gamma^2/3$, $\cL_{\mathrm{conv}}^\star=\gamma^2$, and $\cL_{\CPTP}^\star=0$.}
\label{fig:three_level_diagnostic}
\end{figure}

Figure~\ref{fig:three_level_diagnostic}(a) compares training convergence for three targets, $O_\star\in\{Z,\;0.5Z,\;0.3\id+0.5Z\}$, chosen to lie in $\mathcal{R}_U(Z)$, in $\mathcal{R}_{\mathrm{conv}}(Z)\setminus\mathcal{R}_U(Z)$, and in $\mathcal{R}_{\mathrm{CPTP}}(Z)\setminus\mathcal{R}_{\mathrm{conv}}(Z)$ respectively. The Std-QNN representative reaches zero loss only on $Z$; on $0.5Z$ it converges to the floor $(1-r)^2/3$ imposed by the fixed spectrum, and on $0.3\id+0.5Z$ to a larger floor combining the spectral and identity contributions. The RE-QNN representative reaches zero loss on both trace-zero targets and retains a floor of $\alpha^2$ on $0.3\id+0.5Z$, which is the trace obstruction of the mixed-unitary level. The SQNN representative reaches zero loss on all three, consistent with all three targets lying inside the operator interval $\mathcal{R}_{\mathrm{CPTP}}(Z)$.

Panel (b) displays the analytic optimal-loss landscape over the slice $O_{\alpha,r}=\alpha\id+rZ$ for the three levels. The reachable-observable geometry is visible: the unitary level is restricted to the fixed-spectrum boundary, the mixed-unitary level spans the trace-zero interval, and the CPTP-induced level fills the operator interval bounded by $\alpha+r\le1$ in this quadrant.

Panels (c) and (d) provide quantitative verification of the amplitude-damping witness. For $\gamma\in[0,0.9]$ across five random seeds, the observed loss floors follow the analytic predictions: the Std-QNN representative converges to $\mathcal{L}_U^\star=4\gamma^2/3$, the RE-QNN representative to $\mathcal{L}_{\mathrm{conv}}^\star=\gamma^2$, and the SQNN representative to $\mathcal{L}_{\mathrm{CPTP}}^\star=0$ within numerical precision. The observed separation matches the trace obstruction predicted by Proposition~\ref{prop:local_single_qubit_separation} rather than an optimization artifact.

The agreement between the trained floors and Eq.~\eqref{eq:three_level_loss_floors} across $\gamma\in[0,0.9]$ shows that the three finite parametrizations realize the reachable classes they are intended to represent, and that the separation established in Proposition~\ref{prop:local_single_qubit_separation} appears at the level of an achievable loss rather than only as an algebraic statement. The residual gap at the mixed-unitary level tracks $\gamma^2$ over the whole range, which identifies it as the unreachable identity component rather than an optimization residual.

\subsection{Representative finite-ansatz evidence}
\label{subsec:representative_benchmarks}

Beyond the certified setting of Sec.~\ref{subsec:certified_separation}, we examine how the reachable-observable hierarchy manifests in finite-depth circuits on structured learning tasks. Unlike the loss-floor validation, these experiments have no closed-form optimum: they depend on the specific ans\"atze, the optimizer, the training budget and the ancilla dimension. They are reported as representative finite-ansatz evidence, and are not offered as proofs of representational limitation. In particular, a nonzero floor for a lower-level representative is consistent with the target lying outside that level, but is not by itself a demonstration of it.

\paragraph*{Structured-observable learning.}

Proposition~\ref{prop:local_single_qubit_separation} establishes that for a fixed local readout $M=Z_1$, any observable reachable by unitary or mixed-unitary dynamics must be traceless. A target containing a nonzero identity component therefore lies in $\mathcal{R}_{\mathrm{CPTP}}(M)\setminus\mathcal{R}_{\mathrm{conv}}(M)$ and cannot be exactly realized by mixed-unitary resources.

We construct targets from the structured family
\begin{equation}
    O_{\mathrm{str}}
    =
    c_{\mathbf I}\id
    +
    \sum_{j=1}^{N_{\mathrm{sys}}} c_j Z_j
    +
    \sum_{j<k} c_{jk} Z_j Z_k,
    \qquad c_{\mathbf I}\in[0.15,0.40],
    \label{eq:structured_observable_target}
\end{equation}
which mimics the Heisenberg image of local non-unital channels. The target-construction procedure (App.~\ref{app:structured_obs}) ensures $O_{\mathrm{str}}\in\mathcal{R}_{\mathrm{CPTP}}(M)\setminus\mathcal{R}_{\mathrm{conv}}(M)$. We compare finite-depth representatives of the three hierarchy levels using the empirical MSE in Eq.~\eqref{eq:empirical_mse} and the relative Frobenius distance
\begin{equation}
    D_F
    =
    \frac{\|M_{\mathrm{eff}}-O_{\mathrm{str}}\|_F}
         {\|O_{\mathrm{str}}\|_F},
    \label{eq:frob_distance_def}
\end{equation}
to quantify operator-level agreement. Unless otherwise stated, the RE-QNN employs $K=8$ unitary branches. The SQNN uses $N_{\mathrm{anc}}\in\{1,2,3,4\}$ ancilla qubits across the tested system sizes, with $N_{\mathrm{anc}}\in\{1,2,3\}$ for the $N_{\mathrm{sys}}=3$ traces; scaling resources are specified in App.~\ref{app:numerics_details}.

\begin{figure}[tbp]
    \centering
    \includegraphics[width=0.5\columnwidth]{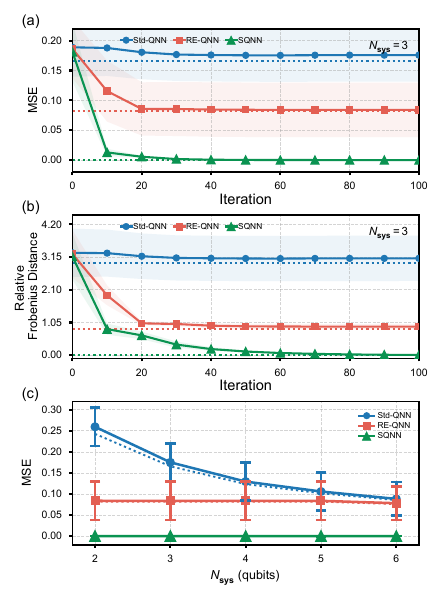}
    \caption{Structured-observable learning with fixed local readout $M=Z_1$. Targets are sampled from Eq.~\eqref{eq:structured_observable_target} with $c_{\mathbf I}\in[0.15,0.40]$.
    (a) Training MSE $\widehat{\mathcal{L}}$ and
    (b) relative Frobenius distance $D_F$ for $N_{\mathrm{sys}}=3$ qubits. Curves show the mean over 10 independently sampled targets and random initializations; shaded regions indicate standard deviation.
    (c) Final MSE versus system size $N_{\mathrm{sys}}\in\{2,3,4,5,6\}$. Markers denote the mean over 10 runs; error bars indicate standard deviation.}
    \label{fig:structured_obs}
\end{figure}

The targets lie outside $\mathcal{R}_{\mathrm{conv}}(M)$ by construction, so a nonzero asymptotic error for the unitary and mixed-unitary representatives is expected. What the experiment probes is whether that expectation survives finite depth, finite parameter count and gradient-based training, and how the residual behaves with system size. The magnitude of the observed floors is not predicted by the theory and should not be read as a measure of the separation.

Figures~\ref{fig:structured_obs}(a) and~\ref{fig:structured_obs}(b) show the learning behavior for $N_{\mathrm{sys}}=3$ qubits. The Std-QNN and RE-QNN representatives converge to persistent nonzero floors in both $\widehat{\cL}$ and $D_F$, consistent with the target lying outside the mixed-unitary reachable class. The SQNN representative reaches lower residual error in both metrics, consistent with the target being realizable within the CPTP-induced class. Figure~\ref{fig:structured_obs}(c) examines $N_{\mathrm{sys}}\in\{2,\dots,6\}$. The SQNN representative maintains near-zero error across the tested sizes. The floors of the Std-QNN and RE-QNN representatives decrease slowly with system size, which is consistent with the shrinking relative weight of the identity component in Eq.~\eqref{eq:structured_observable_target} once the operator norm is held fixed, and does not indicate that the obstruction weakens: the trace constraint $\tr O=0$ holds at every system size.

Within the finite-depth ans\"atze and optimization settings employed here, these results provide representative evidence that the trace obstruction distinguishing the mixed-unitary and CPTP-induced levels can manifest in structured-observable learning tasks. The persistent nonzero floors for the lower-level representatives are consistent with the targets lying outside $\mathcal{R}_{\mathrm{conv}}(M)$ by construction, while the SQNN representative realizes targets within the broader CPTP-induced class.

\paragraph*{Replicated-space R\'enyi-moment learning.}

Subsystem R\'enyi moments $\tr(\rho_A^n)$ can be reformulated as linear observable-learning problems through the replicated-space identity
\begin{equation}
\tr(\rho_A^n) = \tr\left[ V_A^{(n)} \rho_A^{\otimes n} \right],
\label{eq:renyi_identity}
\end{equation}
where $V_A^{(n)}$ is the cyclic shift operator on $n$ replicas of subsystem $A$. For $n>2$, we use the Hermitian part of $V_A^{(n)}$ as the target observable. For the readouts used below the target's membership can be checked directly: the cyclic-shift operator $V_A^{(n)}$ has nonzero trace on the replicated space, whereas the traceless Pauli readout $M=X\otimes I$ generates only traceless elements of $\mathcal{R}_{\mathrm{conv}}(M)$. The Hermitian replicated-space target therefore lies outside the mixed-unitary level, and the hierarchy predicts a persistent floor for the unitary and mixed-unitary representatives.

We test this prediction for the purity ($n=2$) and third-order moment ($n=3$) in two configurations at $N_{\mathrm{sys}}=7$. The first uses a single-qubit subsystem ($N_{\mathrm{sub}}=1$) with fixed readout $M=X\otimes I$. The second uses a larger subsystem ($N_{\mathrm{sub}}=3$) with a random-Hermitian readout $M=H_{\mathrm{rand}}$. We compare finite-depth representatives of the three hierarchy levels using the empirical MSE in Eq.~\eqref{eq:empirical_mse}.

\begin{figure}[tbp]
\centering
\includegraphics[width=0.5\columnwidth]{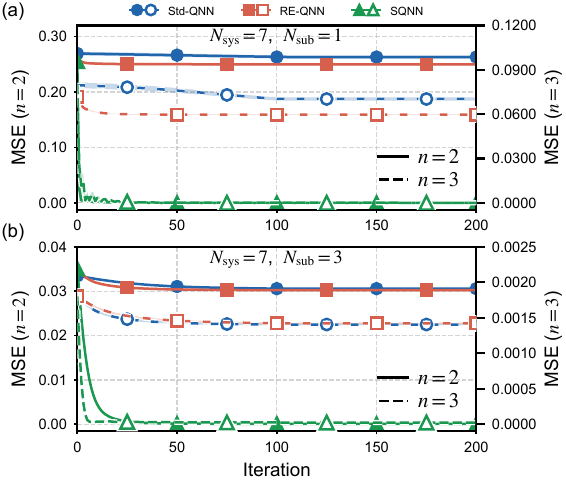}
\caption{R\'enyi-moment learning via replicated-space observable learning. The $n$-th subsystem moment $\tr(\rho_A^n)$ is recast as a linear problem via Eq.~\eqref{eq:renyi_identity}. Curves show mean over 10 independent random initializations; shaded regions indicate standard deviation.
(a) Purity ($n=2$, solid, left axis) and third-order moment ($n=3$, dashed, right axis) for a single-qubit subsystem ($N_{\mathrm{sub}}=1$) in an $N_{\mathrm{sys}}=7$ system with fixed readout $M=X\otimes I$.
(b) Same moments for a larger subsystem ($N_{\mathrm{sub}}=3$) with random-Hermitian readout $M=H_{\mathrm{rand}}$.
}
\label{fig:renyi_training_supporting}
\end{figure}

Figure~\ref{fig:renyi_training_supporting} shows the learning behavior for both configurations. In panel (a), the Std-QNN and RE-QNN representatives converge to persistent nonzero error floors for both the purity ($n=2$, solid) and the third-order moment ($n=3$, dashed), while the SQNN representative reaches near-zero MSE in both cases. Panel (b) shows similar behavior for the larger subsystem with $M=H_{\mathrm{rand}}$: the lower-level representatives retain nonzero floors, while the SQNN representative reaches near-zero error.

Both moment orders and both readout choices show the same pattern: floors for the unitary and mixed-unitary representatives, near-zero error for the SQNN representative. Within the ans\"atze and training budgets used here this is the behaviour the hierarchy predicts for targets outside $\mathcal{R}_{\mathrm{conv}}(M)$, and it indicates that the obstruction identified for a single qubit survives in a seven-qubit replicated-space task. It does not establish that no mixed-unitary circuit of larger depth or branch number would do better, and the residual floors of the SQNN representative reflect finite ancilla dimension.

\subsection{Trainability control}
\label{subsec:trainability_control}

The preceding experiments probe observables designed to lie outside the lower hierarchy levels. As a complementary trainability check, we examine a supervised binary classification task that does not explicitly require observables in $\mathcal{R}_{\mathrm{CPTP}}(M)\setminus\mathcal{R}_{\mathrm{conv}}(M)$. This control experiment confirms that CPTP ans\"atze remain trainable in settings where the hierarchy is not actively probed.

We use the MNIST handwritten-digits dataset~\cite{lecun1998} with binary classification of digit-$3$ versus digit-$8$. Images are encoded into an $N_{\mathrm{sys}}=8$ qubit model using a fixed feature map and fixed Pauli-string readout (App.~\ref{app:classification}). Models are trained by minimizing binary cross-entropy loss.

\begin{figure}[tbp]
\centering
\includegraphics[width=0.5\columnwidth]{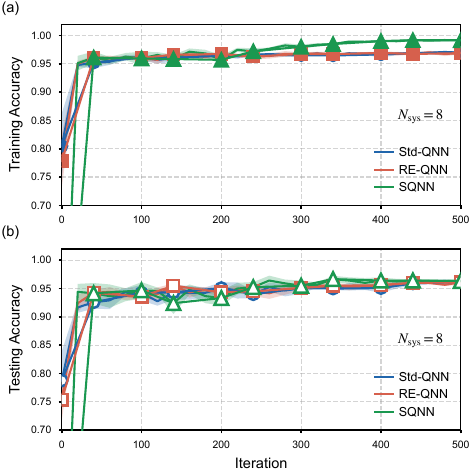}
\caption{MNIST digit-$3$ versus digit-$8$ classification using an $N_{\mathrm{sys}}=8$ qubit model with a fixed Pauli-string readout. (a) Training accuracy and (b) testing accuracy averaged over five independent runs; shaded regions indicate $\pm1$ standard deviation. Additional metrics and visualizations are in App.~\ref{app:classification}.}
\label{fig:clf_accuracy}
\end{figure}

Figures~\ref{fig:clf_accuracy}(a) and~\ref{fig:clf_accuracy}(b) show the training and testing accuracy averaged over five runs. All three architecture representatives converge reliably and achieve comparable final performance, with test accuracies exceeding $96\%$. Additional metrics, confusion matrices, and decision-boundary visualizations are provided in App.~\ref{app:classification}.

The purpose of this benchmark is to rule out a trainability confound, not to test the hierarchy. MNIST digit classification with a fixed Pauli-string readout does not require observables in $\mathcal{R}_{\mathrm{CPTP}}(M)\setminus\mathcal{R}_{\mathrm{conv}}(M)$, so comparable accuracy across the three representatives is the expected outcome and is reported as such. What it establishes is that the ancilla-assisted construction enlarges the reachable class without degrading optimization in a standard supervised setting; taken with Secs.~\ref{subsec:certified_separation} and~\ref{subsec:representative_benchmarks}, it supports reading the floors observed there as reachability constraints rather than as training failures.

\section{Discussion and Limitations}
\label{sec:discussion}

This work characterizes the observable-level expressivity of quantum neural networks under a fixed readout.
When the measurement primitive is constrained by hardware, by tomographic overhead, or by design, the effective observable induced by the trainable dynamics becomes the object that determines the model's representable functions: the hypothesis class is exactly $\mathcal{F}_{\mathcal{R}}$ of Eq.~\eqref{eq:hypothesis_class} for the reachable set $\mathcal{R}$ realized by the available dynamics.

The central results are:
(1) a spectral characterization of each reachable-observable class, a fixed-spectrum orbit, its majorization-described convex hull, and an operator interval;
(2) analytical witnesses proving both inclusions strict for any readout with more than one distinct eigenvalue, detectable on a single qubit at arbitrary system size;
(3) a physical reading of what each resource buys: unitary dynamics preserve the full spectrum of the readout; mixed-unitary ensembles relax this to a majorization constraint while preserving the trace; and CPTP maps, which remain trace-preserving as channels on states, have a unital rather than trace-preserving Heisenberg adjoint, which lifts the trace constraint inherited by mixed-unitary observables and admits non-unital effective observables through environment coupling.

The Stinespring-channelized quantum neural network (SQNN) of Sec.~\ref{sec:qchann_architecture} realizes the CPTP level at the circuit level.
The architecture is not the contribution; it shows that $\mathcal{R}_{\mathrm{CPTP}}(M)$ is accessible with finite ancilla dimension and standard gate sets, and it makes the hierarchy testable in a learning task.

\subsection*{What the Results Do and Do Not Concern}

\textit{Observable expressivity.}
The reachable sets characterize \emph{representational capacity} under a fixed readout: which effective observables, and therefore which linear functionals $\rho\mapsto\tr[O\rho]$, are available.
All separations proved here are statements at this level and at this level only.

\textit{Trainability.}
Nothing here addresses whether gradient-based optimization finds the parameters that realize a reachable observable.
Barren plateaus, landscape structure and initialization are separate questions~\cite{mcclean2018barren,Cerezo2021Cost,Holmes2022Connecting}.
Observable expressivity is necessary but not sufficient for learning success.
The classification control of Sec.~\ref{subsec:trainability_control} checks only that the channelized construction does not degrade optimization in a standard supervised setting; it is not a trainability theory.

\textit{Generalization.}
A larger reachable class does not imply better generalization or sample efficiency.
Enlarging a hypothesis class weakens, rather than strengthens, capacity-based generalization guarantees~\cite{Abbas2021The,Caro2022}.
Sample complexity under the three observable classes is open.

\textit{Ideal CPTP reachable sets versus finite-depth realizations.}
Proposition~\ref{prop:three_level_geometry} describes the set reachable under arbitrary CPTP maps.
A concrete ansatz accesses an architecture-dependent subset limited by depth, connectivity, native gates, noise and ancilla dimension.
Saturation of $\mathcal{R}_{\mathrm{CPTP}}(M)$ requires unrestricted joint unitaries and sufficient ancilla dimension; the finite models used in Sec.~\ref{sec:observable_learning} are representatives of the levels, not exhaustive parametrizations of them.
Characterizing the reachable subset as a function of circuit resources is the natural next step.

\textit{Certified versus representative numerics.}
The results in Sec.~\ref{subsec:certified_separation} are certified: the optimum at each level is known in closed form and the numerics test agreement with it.
The results in Secs.~\ref{subsec:representative_benchmarks} and~\ref{subsec:trainability_control} depend on ansatz, optimizer and budget, and are reported as representative evidence.
Persistent floors there are consistent with the targets lying outside $\mathcal{R}_{\mathrm{conv}}(M)$ by construction, but do not by themselves establish a representational limitation independent of depth and optimization.

\textit{Fixed-readout assumption.}
The separations are fixed-readout statements.
The scalar output is the measured expectation value $f_O(\rho)=\tr[O\rho]$, with no trainable post-processing.
Appending an affine map gives, for normalized states,
\begin{equation}
    a f_O(\rho)+b
    =
    \tr[(aO+b\id)\rho],
    \qquad \tr\rho=1 ,
    \label{eq:affine_effective_observable}
\end{equation}
so affine freedom is equivalent to replacing the effective observable by $aO+b\id$ and can absorb both spectral rescaling and identity shifts.
The amplitude-damping witness $O_\gamma=\gamma\id+(1-\gamma)Z$ lies outside $\mathcal{R}_{\mathrm{conv}}(Z)$ as a fixed-readout observable, but the choice $a=(1-\gamma)^{-1}$, $b=-\gamma(1-\gamma)^{-1}$ maps it into the traceless subspace, so the separation as stated does not survive unrestricted affine post-processing.
The hierarchy characterizes what quantum dynamics can produce \emph{before} classical post-processing is introduced, which is the operationally relevant question when the post-processing budget is itself constrained, and which makes explicit which constraints are dynamical and which are classical.

\textit{Mathematical foundations.}
The majorization characterization of $\mathcal{R}_{\mathrm{conv}}(M)$ rests on Schur--Horn--Uhlmann theory~\cite{nielsen2002majorization,Uhlmann1971,Watrous2018}, and the CPTP characterization on standard Stinespring and Kraus representations~\cite{stinespring1955positive}.
These tools are not new.
What is new here is their use to answer the fixed-readout expressivity question and the resulting resource hierarchy.

\textit{Relation to Existing Approaches.} Several frameworks also move beyond a single unitary circuit while addressing different objects.
Randomness-enhanced and ensemble QNNs~\cite{wu2024randomness} average over unitary readouts and therefore sit at the mixed-unitary level $\mathcal{R}_{\mathrm{conv}}(M)$, subject to majorization and trace constraints.
Noisy, dissipative and reservoir models use open-system dynamics as a learning substrate~\cite{shen2023noisy,Ghasemian2022Dissipative,fujii2017harnessing} but do not characterize the resulting effective observables relative to a fixed readout.
Linear-combination-of-unitaries and block-encoding methods~\cite{childs2012hamiltonian,Gilyen2019Quantum,rall2020quantum,heredge2024non} implement a prescribed non-unitary operator through an enlarged unitary construction, and channel-learning methods identify or approximate an unknown process~\cite{torlai2023process,belov2024channel}.
The question here is the converse: given a fixed measurement and trainable dynamics, which observables become available.

The framework is also complementary to existing expressivity measures rather than competing with them.
Wu \emph{et al.}~\cite{wu2024randomness} identify obstructions arising from limited input dimension, at the level of state-space accessibility.
Fourier analyses of data encoding~\cite{schuld2021effect} characterize the frequency content available to the model.
Effective dimension~\cite{Abbas2021The} quantifies capacity and connects to generalization.
The reachable-observable hierarchy instead characterizes the algebraic structure of the effective observables available under a fixed readout, and can be applied alongside any of these.

\subsection*{Implications and Outlook}

The framework supplies a resource-based criterion for architecture comparison: fixed-readout models can be distinguished by the effective observables they make accessible, rather than by parameter count or depth alone.
Where the measurement primitive is fixed and the structure of the target observable is known or estimable, the hierarchy indicates which resource -- coherence, classical randomness, or environment coupling -- is required, and identifies when additional ancillas cannot help because the target already lies at a lower level.

Open directions include: reachable sets for specific finite-depth ans\"atze; the interaction between observable expressivity and trainability, given that ancilla-assisted models enlarge the reachable set while also enlarging the parameter space; generalization and sample-complexity bounds indexed by observable class; the behaviour of the hierarchy under shot noise and device noise, where the readout itself becomes effectively channelized; and identification of task families whose target observables provably require the CPTP level.

\section{Conclusion}

The framework supplies a resource-based criterion for architecture comparison: fixed-readout models can be distinguished by the effective observables they make accessible, rather than by parameter count or depth alone.
Where the measurement primitive is fixed and the structure of the target observable is known or estimable, the hierarchy indicates which resource -- coherence, classical randomness, or environment coupling -- is required, and identifies when additional ancillas cannot help because the target already lies at a lower level.

Open directions include: reachable sets for specific finite-depth ans\"atze; the interaction between observable expressivity and trainability, given that ancilla-assisted models enlarge the reachable set while also enlarging the parameter space; generalization and sample-complexity bounds indexed by observable class; the behaviour of the hierarchy under shot noise and device noise, where the readout itself becomes effectively channelized; and identification of task families whose target observables provably require the CPTP level.

\section*{Data and Code Availability}
Synthetic data are generated as described in App.~\ref{app:numerics_details}, and MNIST is used for the classification benchmark. The code and scripts needed to reproduce the figures are available from the corresponding author upon reasonable request.

\begin{acknowledgments}
K.L. acknowledges support from the National Natural Science Foundation of China under Grant No. 62571343, the Guangdong Provincial Quantum Science Strategic Initiative (Grant Nos. GDZX2503005, GDZX2403001, GDZX2303001, GDZX2306001), and Guangdong Basic and Applied Basic Research Foundation (No. 2026A1515011661).
\end{acknowledgments}

\appendix

%-------------------------------
\section{Numerical Implementation Details}
\label{app:numerics_details}

This appendix gives the implementation details for the numerical experiments in Sec.~\ref{sec:observable_learning}.
For the certified hierarchy-separation experiments in Sec.~\ref{subsec:certified_separation}, labels are computed from exact noiseless expectation values rather than finite-shot samples.
Unless otherwise stated, each experiment uses $512$ Haar-random training states and $4096$ Haar-random test states.
The states are generated in antithetic pairs: for every one-qubit Haar state $\ket{\psi}$, the orthogonal state $\ket{\psi_\perp}$ is also included.
This removes finite-sample Bloch-vector bias while preserving Haar marginals.
All model classes compared within the same experiment use the same training set, test set, empirical MSE, and optimizer settings.
Table~\ref{tab:resource_accounting} summarizes the resources and training budgets specified for the numerical benchmarks.

\begin{table*}[!t]
\centering
\caption{Resource and training summary for the numerical benchmarks. The table is intended to make the representative comparisons transparent; the finite-ansatz benchmarks are not parameter-matched expressivity theorems. N/A indicates not applicable, and the Kraus-rank column gives the architectural upper bound when applicable.}
\label{tab:resource_accounting}
\resizebox{\textwidth}{!}{%
\begin{tabular}{lcccccll}
\hline
\hline
Benchmark & Model & $N_{\rm sys}$ & $N_{\rm anc}$ & $K$ & Kraus rank & Optimizer/budget & Samples and interpretation \\
\hline
Three-level diagnostic & Std-QNN & 1 & 0 & N/A & 1 & L-BFGS-B trajectory & 512 train, 4096 test; unitary-orbit representative \\
Three-level diagnostic & RE-QNN & 1 & 1 branch qubit & 2 & $\le 2$ & L-BFGS-B trajectory & 512 train, 4096 test; mixed-unitary representative \\
Three-level diagnostic & SQNN & 1 & 2 & N/A & $\le 4$ & L-BFGS-B trajectory & 512 train, 4096 test; CPTP representative for the plotted slice \\
Amplitude damping & Std-QNN & 1 & 0 & N/A & 1 & L-BFGS-B, 3 restarts & 512 train, 4096 test; five seeds over $\gamma=0,\ldots,0.9$ \\
Amplitude damping & RE-QNN & 1 & 0 & 4 & $\le 4$ & L-BFGS-B, 3 restarts & Same datasets; mixed-unitary loss-floor baseline \\
Amplitude damping & SQNN & 1 & 1 & N/A & $\le 2$ & L-BFGS-B, 3 restarts & Same datasets; amplitude-damping Stinespring family \\
Structured observable & Std-QNN & 2--6 & 0 & N/A & 1 & Adam, 1000 epochs & $N_D=200$--$800$; fixed local readout \\
Structured observable & RE-QNN & 2--6 & 0 & 8 & $\le 8$ & Adam, 1000 epochs & Same datasets; mixed-unitary baseline \\
Structured observable & SQNN & 2--6 & 1--4 & N/A & $\le 2^{N_{\rm anc}}$ & Adam, 1000 epochs & Same datasets; $N_{\rm sys}=3$ traces use $N_{\rm anc}=1,2,3$ \\
R\'enyi-moment & Std-QNN & 7 & 0 & N/A & 1 & 1500 epochs; rates in Table~\ref{tab:renyi_configs} & $N_D=600$ for $N_{\rm sub}=1$, $300$ for $N_{\rm sub}=3$ \\
R\'enyi-moment & RE-QNN & 7 & 0 & 4 or 6 & $\le K$ & 1500 epochs; rates in Table~\ref{tab:renyi_configs} & $K=4$ for $N_{\rm sub}=1$, $K=6$ for $N_{\rm sub}=3$ \\
R\'enyi-moment & SQNN & 7 & 2 or 3 & N/A & $\le 2^{N_{\rm anc}}$ & 1500 epochs; rates in Table~\ref{tab:renyi_configs} & $N_{\rm anc}=2$ for $N_{\rm sub}=1$, $3$ for $N_{\rm sub}=3$ \\
Classification & Std-QNN & 8 & 0 & N/A & 1 & Adam, 500 epochs, $B=64$ & 480 train, 120 test; trainability check \\
Classification & RE-QNN & 8 & 0 & 4 & $\le 4$ & Adam, 500 epochs, $B=64$ & Same split; trainable mixture weights \\
Classification & SQNN & 8 & 2 & N/A & $\le 4$ & Adam, 500 epochs, $B=64$ & Same split; comparable classification performance \\
\hline
\hline
\end{tabular}%
}
\end{table*}

\subsection{Three-level diagnostic details}

To verify the analytical predictions for reachable observables across the three hierarchy levels, we construct minimal parametric circuits that realize each class and numerically confirm that their optimization landscapes reproduce the theoretical loss floors.

The diagnostic uses target observables in the one-qubit slice
\begin{equation}
    O_{\alpha,r}=\alpha\id+rZ,
    \qquad \alpha,r\in[0,1],
\end{equation}
with empirical loss given by Eq.~\eqref{eq:empirical_mse} over Haar-random input states.
The three representative targets in Fig.~\ref{fig:three_level_diagnostic}(a),
\begin{equation}
    O_\star=Z,
    \qquad
    O_\star=0.5Z,
    \qquad
    O_\star=0.3\id+0.5Z ,
    \label{eq:three_level_targets}
\end{equation}
test three distinct reachability regimes: unitary reachability, interior of the mixed-unitary set, and violation of the mixed-unitary trace constraint, respectively.

For the unitary hierarchy level, we parametrize the effective observable using a single-qubit rotation,
\begin{equation}
    U(\theta_U)=R_y(-\theta_U),
    \
    M_{\mathrm{eff}}^{U}(\theta_U)
    =
    U^\dagger(\theta_U)ZU(\theta_U)
    =
    \sin\theta_U\,X+\cos\theta_U\,Z .
\end{equation}
This one-parameter family suffices to explore the unitary orbit intersecting the $\{\id,Z\}$ plane.

To realize the mixed-unitary observable class, we implement an explicit probabilistic mixture via an ancilla-controlled branch.
An ancilla qubit $b$ prepared as
\begin{equation}
    R_y(2\eta)\ket{0}_b
    =
    \sqrt{p}\ket{0}_b+
    \sqrt{1-p}\ket{1}_b,
    \qquad p=\cos^2\eta ,
\end{equation}
controls a system Pauli-$X$ gate, which is then discarded.
The resulting system channel,
\begin{equation}
    \Phi_p(\rho)=p\rho+(1-p)X\rho X,
\end{equation}
induces the effective observable
\begin{equation}
    M_{\mathrm{eff}}^{\mathrm{conv}}(p)
    =
    \Phi_p^\dagger(Z)
    =
    (2p-1)Z,
    \qquad p\in[0,1] .
\end{equation}
This parametrization sweeps the unitary-convex segment $[-Z,Z]$.

The full CPTP-induced class is realized by composing the mixed-unitary contraction with an amplitude-damping channel.
A second ancilla qubit $a$ mediates the Stinespring dilation of the damping channel $\cE_{\alpha'}^{\mathrm{AD}}$, where $\alpha'$ controls the damping strength.
The composite channel,
\begin{equation}
    \cE_{\alpha',p}=\cE_{\alpha'}^{\mathrm{AD}}\circ\Phi_p,
\end{equation}
generates the effective observable
\begin{equation}
    M_{\mathrm{eff}}^{\CPTP}(\alpha',p)
    =
    \alpha'\id+(1-\alpha')(2p-1)Z,
    \
    \alpha'\in[0,1],\ p\in[0,1] .
\end{equation}
This two-parameter family spans the quadrilateral operator interval with $\alpha,r\in[0,1]$ and $\alpha+r\le1$ shown in Fig.~\ref{fig:three_level_diagnostic}.
The amplitude-damping block serves as a controlled non-unital channel realization, enabling the identity shift beyond mixed-unitary constraints.

For each hierarchy level, we optimize the parametric effective observable to match each target using L-BFGS-B over the natural parameter domains ($\theta_U,\eta\in[0,\pi/2]$, and additionally $\theta_{\mathrm{AD}}\in[0,\pi/2]$ where $\alpha'=\sin^2\theta_{\mathrm{AD}}$ for the CPTP case).
Figure~\ref{fig:three_level_diagnostic}(a) shows the loss evolution during optimization, demonstrating convergence to the analytically predicted floors.

The analytical loss floors arise from the geometry of each reachable set.
For a target $O_{\alpha,r}=\alpha\id+rZ$ and a traceless effective observable $E=\bm v\cdot\bm\sigma$, the Haar-averaged loss is
\begin{equation}
    \mathcal L(E;O_{\alpha,r})
    =
    \alpha^2+\frac{\|r\hat{\bm z}-\bm v\|_2^2}{3}.
\end{equation}
Minimizing over each reachable observable class yields
\begin{equation}
    \mathcal L_U^\star(\alpha,r)
    =
    \alpha^2+\frac{(1-r)^2}{3},
    \qquad
    \mathcal L_{\mathrm{conv}}^\star(\alpha,r)
    =
    \alpha^2,
\end{equation}
for the unitary and mixed-unitary levels, while
\begin{equation}
    \mathcal L_{\CPTP}^\star(\alpha,r)=0
\end{equation}
inside the CPTP-induced operator interval ($\alpha,r\in[0,1]$, $\alpha+r\le1$).
The landscape plots in Fig.~\ref{fig:three_level_diagnostic}(b) show these theoretical optima over the target space, and the numerical results in panel (a) confirm convergence to these values within optimization tolerance.
This agreement validates the analytical characterization of reachable observables across all three hierarchy levels.

\subsection{Amplitude-Damping Circuit Details}
\label{app:amp_circuit}

To verify the predicted reachable-set boundaries for non-unital CPTP dynamics (Sec.~\ref{sec:reachability}), we implement a single-qubit amplitude-damping channel as the minimal example of a quantum channel that cannot be realized by unitary conjugation alone.
The channel is constructed via the Stinespring dilation with a one-qubit ancilla initialized in $\ket{0}_a$, where subscripts $s$ and $a$ denote system and ancilla qubits.

The Stinespring unitary acts as
\begin{align}
    \ket{0}_s\ket{0}_a &\mapsto \ket{0}_s\ket{0}_a,\\
    \ket{1}_s\ket{0}_a &\mapsto
    \sqrt{1-\gamma}\ket{1}_s\ket{0}_a
    +
    \sqrt{\gamma}\ket{0}_s\ket{1}_a.
\end{align}
Tracing out the ancilla yields the Kraus operators in Eq.~\eqref{eq:amplitude_damping_kraus}.
Equivalently, the Stinespring unitary may be implemented as an excitation-preserving two-qubit rotation in the subspace spanned by $\{\ket{1}_s\ket{0}_a,\ket{0}_s\ket{1}_a\}$ with rotation angle $\theta_\gamma=\arcsin\sqrt{\gamma}$, together with identity action on $\ket{0}_s\ket{0}_a$.

In the numerical experiments of Sec.~\ref{sec:observable_learning}, the SQNN ansatz optimizes the parameter $\gamma'=\sin^2\theta$ directly, allowing continuous exploration of the one-parameter family of amplitude-damping channels.
At the level of effective observables $\mathcal{O}_{\text{eff}}(\rho_0) = \text{Tr}[M\mathcal{E}(\rho_0)]$ for fixed readout $M=Z_s$, varying $\gamma' \in [0,1]$ sweeps through the CPTP-reachable set for non-unital dynamics.
The circuit construction ensures that $\gamma'=0$ recovers the identity channel (unitary-reachable limit), while $\gamma'>0$ probes the expansion of the reachable set beyond unitary conjugation.
This controlled parameterization enables direct verification of the loss floor predictions for target observables that lie outside the unitary-reachable set.

\subsection{Structured-observable learning benchmark details}
\label{app:structured_obs}

We verify the analytical predictions of Sec.~\ref{sec:reachability} by training representative finite-dimensional implementations from each hierarchy level on a structured observable learning task. The goal is to demonstrate that the expressivity gaps predicted by the reachable observable characterizations manifest in finite-ansatz performance.

To probe the hierarchy between unitary, mixed-unitary, and CPTP-induced observables, we sample targets from the structured family
\begin{equation}
O_{\mathrm{str}}
=
c_{\mathbf I}\id
+
\sum_j c_j Z_j
+
\sum_{j<k} c_{jk} Z_j Z_k,
\end{equation}
where coefficients are drawn uniformly as
$c_{\mathbf I}\in[0.15,0.40]$,
$c_j\in[-0.15,0.15]$,
and
$c_{jk}\in[-0.10,0.10]$,
then rescaled to satisfy $\|O_{\mathrm{str}}\|_\infty\le1$.
This family generates targets with non-trivial two-body correlations that lie outside the unitary orbit of single-qubit diagonal readouts for sufficiently large system sizes, enabling direct testing of expressivity constraints.

For each target, training data consists of $N_D$ Haar-random pure input states $\{\ket{\psi_m}\}_{m=1}^{N_D}$ with exact labels $T_m=\bra{\psi_m}O_{\mathrm{str}}\ket{\psi_m}$.
Models are trained to minimize the empirical MSE of Eq.~\eqref{eq:empirical_mse}, and operator-level agreement is quantified via the relative Frobenius distance of Eq.~\eqref{eq:frob_distance_def}.

We realize the three reachable observable classes using the following finite-dimensional parametrizations. All unitaries are implemented as $U(\theta)=e^{-iH(\theta)}$ with Hermitian generators $H(\theta)=A(\theta)+A^\dagger(\theta)$ initialized with small random complex entries.

\textit{Unitary level:} The effective observable is a standard unitary conjugation,
\begin{equation}
    M_{\mathrm{eff}}^{\mathrm{U}}(\theta) = U^\dagger(\theta) M U(\theta), \quad U \in \mathcal{U}(2^{N_{\mathrm{sys}}}),
\end{equation}
corresponding to the reachable set $\mathcal{R}_{U}(M)$ of Eq.~\eqref{eq:unitary_orbit_set}.

\textit{Convex-unitary level:} A finite mixture of $K=8$ unitary branches represents the convex hull,
\begin{equation}
    M_{\mathrm{eff}}^{\mathrm{conv}}(\theta, \phi) = \sum_{i=1}^{K} p_i(\phi) U_i^\dagger(\theta_i) M U_i(\theta_i),
\end{equation}
where mixing probabilities $p_i(\phi)$ are softmax-parametrized via learned logits $\phi_i$, with independent unitaries $U_i(\theta_i)$.
This approximates the mixed-unitary reachable set $\mathcal{R}_{\mathrm{conv}}(M)$ of Eq.~\eqref{eq:conv_set}.

\textit{CPTP level:} A Stinespring dilation with $N_{\mathrm{anc}}$ ancilla qubits initialized in $\ket{0}_a^{\otimes N_{\mathrm{anc}}}$ implements the dual channel via
\begin{equation}
    M_{\mathrm{eff}}^{\mathrm{CPTP}}(\theta) = \cE_\theta^\dagger(M) = \sum_{\mu=0}^{2^{N_{\mathrm{anc}}}-1} K_\mu^\dagger(\theta) M K_\mu(\theta),
\end{equation}
where Kraus operators $K_\mu(\theta) = {}_a\bra{\mu} U_{sa}(\theta) \ket{0}_a$ are recovered via partial trace over the ancilla basis $\{\ket{\mu}_a\}$.
This construction generates a subset of the full CPTP-induced reachable set $\mathcal{R}_{\mathrm{CPTP}}(M)$ of Eq.~\eqref{eq:cptp_set}, with expressivity controlled by $N_{\mathrm{anc}}$.

All models are optimized for $1000$ epochs using the Adam optimizer with learning rate $\eta = 10^{-2}$.
For the scaling analysis in Fig.~\ref{fig:structured_obs}(c), the training set size grows from $N_D=200$ to $N_D=800$ as system size increases from $N_{\mathrm{sys}}=2$ to $N_{\mathrm{sys}}=6$.
The SQNN utilizes ancilla counts $N_{\mathrm{anc}} \in \{1,2,3,4\}$ across tested system sizes, providing a controlled probe of finite-resource CPTP expressivity.
For each configuration, results are averaged over 5 independent random initializations, with error bars indicating standard error of the mean.

\subsection{Replicated-space R\'enyi-moment benchmark details}
\label{app:renyi_details}

To verify the analytical characterization of reachable observables developed in Sec.~\ref{sec:theory}, we implement finite-depth ansätze for each dynamical class and evaluate their ability to approximate target effective observables. We focus on the replicated-space cyclic-shift observable discussed below Eq.~\eqref{eq:renyi_identity}, whose Hermitian representative encodes subsystem R\'enyi moments $\tr[(\rho_A)^n]$.

Each model is optimized to minimize the empirical mean-squared error
\begin{equation}
    \hat{\mathcal{L}} = \frac{1}{N_D} \sum_{m=1}^{N_D} \left( y_m - \tr[M_{\mathrm{eff}} \rho_{\mathrm{rep}}^{(m)}] \right)^2,
    \label{eq:renyi_app_mse}
\end{equation}
where the target labels $y_m = \tr[(\rho_A^{(m)})^n]$ are computed exactly from Haar-random pure states on the $N_{\mathrm{sys}}$-qubit global system, and $\rho_{\mathrm{rep}}^{(m)}$ denotes the corresponding replicated input state.

We consider four representative configurations summarized in Table~\ref{tab:renyi_configs}: two subsystem dimensions ($N_{\mathrm{sub}} \in \{1,3\}$), two moment orders ($n \in \{2,3\}$), and two fixed readouts—the traceless Pauli $M = X \otimes I$ for $N_{\mathrm{sub}}=1$ and a normalized random Hermitian $M = H_{\mathrm{rand}}$ for $N_{\mathrm{sub}}=3$. In all cases the total system size is fixed at $N_{\mathrm{sys}} = 7$ qubits. The random readout $H_{\mathrm{rand}}$ is drawn once from a Gaussian orthogonal ensemble and normalized to unit operator norm.

\begin{table}[tbp]
\centering
\caption{Replicated-space R\'enyi-moment verification configurations. Results are averaged over 10 independent initializations. Std-QNN, RE-QNN, and SQNN denote unitary, mixed-unitary, and CPTP-realizable dynamics, respectively. Ancilla count $N_{\rm anc}$ and layer depth $K$ determine finite-ansatz expressivity. All models are trained using Adam with batch gradient descent until loss stabilization (typically ${\sim}1500$ epochs; learning rates $\eta$ as shown).}
\label{tab:renyi_configs}
\begin{tabular}{cccccccc}
\hline
$N_{\rm sub}$ & $n$ & Readout & $N_D$ & $N_{\rm anc}$ & $K$ & $\eta$ & Model classes \\
\hline
1 & 2 & $X\otimes I$ & 600 & 2 & 4 & $5{\times}10^{-4}$ & Std/RE/SQNN \\
1 & 3 & $X\otimes I$ & 600 & 2 & 4 & $1{\times}10^{-3}$ & Std/RE/SQNN \\
3 & 2 & $H_{\rm rand}$ & 300 & 3 & 6 & $1{\times}10^{-2}$ & Std/RE/SQNN \\
3 & 3 & $H_{\rm rand}$ & 300 & 3 & 6 & $1{\times}10^{-2}$ & Std/RE/SQNN \\
\hline
\end{tabular}
\end{table}

The three ansatz classes—Std-QNN (unitary), RE-QNN (mixed-unitary via random reset), and SQNN (CPTP-realizable via Stinespring dilation)—correspond to the three reachable-observable hierarchies characterized analytically in Sec.~\ref{sec:theory}. Each uses the same layer structure as the observable-learning verification (Sec.~\ref{subsec:representative_benchmarks}), adapted to the replicated-space input dimension. Parameters are initialized uniformly at random over $[0, 2\pi)$ and optimized using the Adam optimizer with full-batch gradient descent.

The results in Fig.~\ref{fig:renyi_training_supporting} are consistent with the reachability characterization of Proposition~\ref{prop:three_level_geometry}: the CPTP-level representative reaches lower loss floors than the unitary and mixed-unitary representatives on targets constructed to lie outside $\mathcal{R}_{\mathrm{conv}}(M)$. The residual floors present in all models reflect finite ansatz depth $K$ and finite ancilla dimension $N_{\rm anc}$, and the comparison is between representatives of the three levels rather than between optimized architectures.

\subsection{Classification Benchmark Details}
\label{app:classification}

This appendix provides the implementation details for the supervised learning benchmark presented in Sec.~\ref{subsec:trainability_control}.

We utilize the MNIST handwritten-digits dataset, restricted to a binary classification between digit-3 and digit-8. We draw 300 samples per class, for a total dataset of $N_D=600$ samples. The data is partitioned into a stratified 80/20 split, yielding $N_{\mathrm{train}} = 480$ and $N_{\mathrm{test}} = 120$. Each $28\times 28$ greyscale image is flattened and compressed via Principal Component Analysis (PCA) to $N_{\mathrm{sys}} = 8$ components, which retains approximately $80\%$ of the explained variance. These components are rescaled to the interval $[0, \pi]$ using a min-max scaler to suit the periodic nature of quantum gates.

The compressed feature vector $\mathbf{x} \in [0, \pi]^8$ is encoded into an 8-qubit quantum state $|\psi(\mathbf{x})\rangle$ via a shallow parametric circuit:
\begin{equation*}
|\psi(\mathbf{x})\rangle = \left( \bigotimes_{i=1}^8 R_Y(x_i) \right) \left( \bigotimes_{i=1}^3 R_Z(x_i) \right) |0\rangle^{\otimes 8},
\end{equation*}
where $R_Y(\theta) = \exp(-i \theta Y / 2)$ and $R_Z(\theta) = \exp(-i \theta Z / 2)$.

All three model classes Std-QNN, RE-QNN, and SQNN are trained to minimize the binary cross-entropy (BCE) loss:
\begin{align*}
\mathcal{L}
&=
-\frac{1}{B}\sum_{m=1}^{B}\ell_m,
\nonumber\\
\ell_m
&=
y_m \log\!\big(\sigma(f_m)\big)
+
(1-y_m)\log\!\big(1-\sigma(f_m)\big),
\end{align*}
where $B$ denotes the mini-batch size,
$f_m = \langle \psi(\mathbf{x}_m)| M_{\mathrm{eff}}
|\psi(\mathbf{x}_m)\rangle$ is the model output for the
$m$-th sample, $\sigma(z)=(1+e^{-z})^{-1}$ is the sigmoid
activation, and $y_m\in\{0,1\}$ is the ground-truth label.

The Std-QNN employs a single trainable unitary on the system register. The RE-QNN uses $K=4$ unitary branches with trainable mixture weights, while the SQNN employs $N_{\mathrm{anc}}=2$ ancilla qubits and the induced channel observable of Eq.~\eqref{eq:cptp_heisenberg_observable}. All models are optimized using Adam with mini-batch size $B=64$ for 500 epochs.
\begin{table}[tbp]
\centering
\caption{MNIST digit-$3$ versus digit-$8$ classification results for $N_{\mathrm{sys}}=8$. Values represent the average $\pm$ one standard deviation over five independent runs.}
\label{tab:classification_results}
\renewcommand{\arraystretch}{1.2}
\resizebox{\columnwidth}{!}{%
\begin{tabular}{lcccc}
\hline\hline
\textbf{Model} & \textbf{Test Acc.} & \textbf{Precision} & \textbf{Recall} & \textbf{F1} \\
\hline
Std-QNN & $0.963 \pm 0.002$ & $0.971 \pm 0.007$ & $0.952 \pm 0.007$ & $0.961 \pm 0.002$ \\
RE-QNN  & $0.967 \pm 0.005$ & $0.969 \pm 0.007$ & $0.952 \pm 0.004$ & $0.960 \pm 0.005$ \\
SQNN    & $0.972 \pm 0.005$ & $0.968 \pm 0.004$ & $0.958 \pm 0.008$ & $0.963 \pm 0.005$ \\
\hline
\end{tabular}%
}
\end{table}

\begin{figure}[htbp]
\centering
\includegraphics[width=0.5\columnwidth]{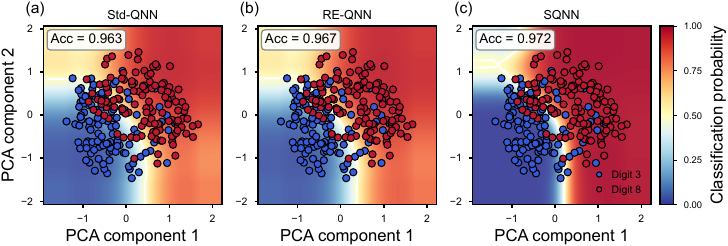}
\caption{
Decision-boundary visualizations projected onto the first two PCA components for the (a) Std-QNN, (b) RE-QNN, and (c) SQNN. Background colors indicate the predicted classification probability. The displayed boundaries correspond to the best-performing run of each architecture.
}
\label{fig:classification_appendix}
\end{figure}

Decision-boundary visualizations are shown in Fig.~\ref{fig:classification_appendix}. The boundaries are generated by evaluating the trained models on a dense grid in the first two PCA coordinates. The grid points are mapped back to the original feature space using the inverse PCA transform and encoded into quantum states. The displayed probabilities correspond to the best-performing run of each architecture.

\bibliography{apssamp}

\end{document}